\documentclass[twocolumn,preprintnumbers,pra,amsmath,amssymb,superscriptaddress,showpacs]{revtex4}
\usepackage{epsfig}
\usepackage{graphicx}

\begin{document}
\title{Reduced randomness in quantum cryptography\\ 
with sequences of qubits encoded in the same basis}

\author{L.-P. Lamoureux}
\affiliation{Quantum Information and Communication, Ecole
Polytechnique, CP 165, Universit\'e Libre de Bruxelles, 1050
Brussels, Belgium}

\author{H. Bechmann-Pasquinucci}
\affiliation{Quantum Information Theory group (QUIT), Dipartimento di Fisica
``A.\,Volta'' and INFM - Unit\`a di Pavia, Via Bassi 6, I--27100  Pavia, Italy}
\affiliation{UCCI.IT, via Olmo 26, I-23888 Rovagnate, Italy}

\author{N. J. Cerf}
\affiliation{Quantum Information and Communication, Ecole
Polytechnique, CP 165, Universit\'e Libre de Bruxelles, 1050
Brussels, Belgium}

\author{N. Gisin}
\affiliation{Group of Applied Physics, University of Geneva, 1211 Geneva 4, Switzerland}

\author{C. Macchiavello}
\affiliation{Quantum Information Theory group (QUIT), Dipartimento di Fisica
``A.\,Volta'' and INFM - Unit\`a di Pavia, Via Bassi 6, I--27100  Pavia, Italy}

\begin{abstract}
We consider the cloning of sequences of qubits prepared in the states
used in the BB84 or 6-state quantum cryptography protocol, and show 
that the single-qubit fidelity is unaffected even if entire sequences 
of qubits are prepared in the same basis. 
This result is of great importance for practical quantum cryptosystems 
because it reduces the need for high-speed random number generation
without impairing on the security against finite-size attacks.
\end{abstract}
\pacs{03.67.-a,03.65.-w}
\maketitle

\section{INTRODUCTION}
The security of quantum cryptography \cite{BB84,GRTZ,Ekert} is based on
two main ingredients.  The first refers to the impossibility of perfectly
cloning some unknown quantum state selected from a nonorthogonal set
\cite{WZ}. As a result, the potential eavesdropper Eve cannot clone the quantum
state transmitted by Alice and re-transmit it undisturbed to the receiver Bob. 
The second ingredient, although often mentioned only implicitly 
in the litterature, is also an absolute requirement:
truly random numbers must be available on both Alice's and Bob's sides.
Indeed, with pseudo-random number generators, the sequence of choices
made by Alice and Bob could in principle be predicted by Eve if the seed
is known to her. Clearly, quantum cryptography should use quantum randomness.  
But, in practice, this is a severe constraint because a complete protocol requires a huge amount of random numbers, from Alice's state choices to Bob's basis choices, as well as for the random choices and random permutations 
needed in error correction and privacy amplification. Making high-speed
quantum random-number generators is a big technological challenge, so that
most realizations of quantum cryptography today rely 
on an active \footnote{Note that if the choice is passive, based on quantum
effects (e.g. the photon being detected at one or the other output port of
a beam splitter), then it is still not completely equivalent to using
a quantum random-number generator. Indeed, in the latter case, 
the photon involved in the random-number generation is generated locally,
and has not been transmitted over the line and potentially tapped by Eve.}
choice that uses a standard random-number generator.
It is therefore of a great importance to investigate whether this requirement
of high-rate random number generation can be relaxed, at least in part.

In this paper, we consider a variant of the BB84 \cite{BB84} or six-state \cite{dagmar,BG} protocols in which the basis chosen for encoding
is kept unchanged over long sequences of qubits instead of being drawn
at random for each qubit. We show that, quite surprisingly, the security 
is unaffected by this modification of the protocol 
although the random number generation rate is significantly reduced.
The BB84 and six-state protocols are amongst the cryptographic
schemes for which the security has exhaustively been studied. In 
various cases the optimal eavesdropping strategy has been found explicitly  
\cite{griffiths,dagmar,BG}, and was shown to coincide 
with approximate cloning \cite{CBKG}. For this reason, we restrict our
analysis to cloning-based attacks in the following.

We consider the cloning of sequences of $N$ qubits. In each 
sequence the qubits are prepared in the same basis, but the state is chosen 
at random among the basis states \footnote{Note that 
this situation is different from 
the usual scenario of $N\to M$ cloning transformations, where $N$ identical
replicas of a quantum state are considered at the input, and are used
to produce $M$ clones.}.  This is viewed as the optimal eavesdropping
attack against a quantum cryptographic protocol in which
we do not restrict Alice and Bob to make random choices of bases 
for every qubit, but allow them to use the same basis for the
entire length-$N$ sequence ($N$ is assumed to be
publicly known).  That is, for each sequence, Alice and Bob make new and
independent random choices of bases. At first sight, one could imagine that
this encoding would increase Eve's knowledge about the secret key, but we
shall see that for the class of cloning transformations we have studied,
this is not the case:  Eve's optimal cloning attack provides her
with no more Shannon information, for a given quantum bit error rate,
than in the usual case where Alice and Bob make random basis-choices 
for each qubit and Eve applies a cloning attack on each qubit.
Under the assumption that this class of approximate cloning transformations
corresponds to the optimal eavesdropping 
strategy, we have thus proven that the requirement for random number 
generation can be reduced without impairing on the security 
against finite-size attacks \cite{gisin}.

The paper is organized as follows.  In Section II, we describe a
general formalism for quantum cloning \cite{cerfansatz1,cerfansatz2},
and adapt it to the case of interest here.  
In Sections III and IV, we apply this formalism to 2-qubit cloning attacks 
in the BB84 and six-state protocols, respectively,
and show that using the same bases does not affect the cloning fidelity.
Section V contains a generalization of these results in dimensions being
any power of two. Finally, in Section VI, we summarize our results.

\section{General quantum cloning formalism}
We refer to a general class of cloning transformations as defined 
in Refs.~\cite{cerfansatz1,cerfansatz2}. 
Considering an arbitrary state $|\psi\rangle$ in a $2^N$-dimensional Hilbert 
space, we wish to produce two (approximate) clones.  
The class of cloning transformations we will analyze is built following the 
"Cerf ansatz": if the input state is $|\psi\rangle$, then the resulting joint 
state of the two output clones (noted $E$ and $B$) and the cloning machine 
(noted $C$) is:
\begin{eqnarray}
|\psi\rangle&\rightarrow& \sum^{2^N-1}_{\overset{-}{m},\overline{n}=0}a_{\overline{m},\overline{n}}U_{\overline{m},\overline{n}}|\psi\rangle_E|B_{\overline{m},\overline{n}}\rangle_{B,C}\nonumber\\
&=&\sum^{2^N-1}_{\overline{m},\overline{n}=0}b_{\overline{m},\overline{n}}U_{\overline{m},\overline{n}}|\psi\rangle_B|B_{\overline{m},\overline{n}}\rangle_{E,C} \label{ansatz}
\end{eqnarray}
where the couple $\{\overline{m},\overline{n}\}\Leftrightarrow\{m_1\ldots m_N,n_1\ldots n_N\}$ and $m_i, n_i \in \{0,1\}$.
Here, $E$, $B$ and $C$ are $2^N$-dimensional systems and $U_{\overline{m},\overline{n}}$ is defined as  
\begin{equation}
U_{\overline{m},\overline{n}} =\bigotimes^N_{i=1}X^{m_i}Z^{n_i}.
\end{equation}
where $X^{m_i}Z^{n_i}$ represents the identity and the three Pauli matrices
\begin{eqnarray}
X^0Z^0 &=& I \nonumber \\
X^1Z^0 &=& \sigma_x \nonumber \\
X^0Z^1 &=& \sigma_z \nonumber \\
X^1Z^1 &=&-i\sigma_y. \nonumber
\end{eqnarray}
Here, $|B_{\overline{m},\overline{n}}\rangle$ is defined as
\begin{equation}
|B_{\overline{m},\overline{n}}\rangle = \sum_{\overline{k}=0}^{2^N-1}(-1)^{(\overline{k}\cdot\overline{n})}|\overline{k}\rangle|\overline{k}+\overline{m}\rangle
\label{bell}
\end{equation}
where $\overline{k}\cdot \overline{n}$ represents the bitwise scalar product, i.e. $\overline{k}\cdot \overline{n} = \sum_{i}k_in_i$. Thus,
$U_{\overline{m},\overline{n}}$ is the tensor product of $N$ Pauli matrices each acting on a two-dimensional subsystem.  
An $error$ operator $U_{m_i,n_i}$ is associated  to each subsystem.
Such an operator shifts the state by $m_i$ units (modulo $2$) in the computational basis, and multiplies it by a phase so as to shift its Fourier transform by $n_i$ units (modulo $2$). Eq.~(\ref{bell}) defines the $d^2$ generalized Bell states for a pair of $2^N$-dimensional systems with
$|B_{\overline{m},\overline{n}}\rangle = U_{\overline{m},\overline{n}}\otimes I \ |B_{\overline{0},\overline{0}}\rangle$.

Tracing over systems $B$ and $C$ (or $E$ and $C$) yields the final states of clone $E$ (or clone $B$):  if the input state is $|\psi\rangle$, the clones $E$ and $B$ are in a mixture of the states $|\psi_{\overline{m},\overline{n}}\rangle = U_{\overline{m},\overline{n}}|\psi\rangle$ with respective weights $p_{\overline{m},\overline{n}}$ and $q_{\overline{m},\overline{n}}$:
\begin{eqnarray}
\rho_E &=& \sum_{\overline{m},\overline{n}=0}^{2^N-1}p_{\overline{m},\overline{n}}|\psi_{\overline{m},\overline{n}}\rangle\langle\psi_{\overline{m},\overline{n}}|\nonumber\\
\rho_B &=& \sum_{\overline{m},\overline{n}=0}^{2^N-1}q_{\overline{m},\overline{n}}|\psi_{\overline{m},\overline{n}}\rangle\langle\psi_{\overline{m},\overline{n}}|
\label{output}
\end{eqnarray}
In addition, the weight functions of the two clones ($p_{\overline{m},\overline{n}}$ and $q_{\overline{m},\overline{n}}$) are related by
\begin{equation}
p_{\overline{m},\overline{n}} = |a_{\overline{m},\overline{n}}|^2,\quad q_{\overline{m},\overline{n}} = |b_{\overline{m},\overline{n}}|^2,
\end{equation}
where $a_{\overline{m},\overline{n}}$ and $b_{\overline{m},\overline{n}}$ are two (complex) amplitude functions that are dual under $N$ two-dimensional Fourier transforms:
\begin{equation}
b_{\overline{m},\overline{n}} = \frac{1}{2^N}\sum_{\overline{x},\overline{y}=0}^{2^N-1}(-1)^{\overline{n}\cdot \overline{x}-\overline{m}\cdot\overline{y}}a_{\overline{x},\overline{y}}.
\label{ft}
\end{equation}
The fidelity of a clone, say $E$, is given by
\begin{equation}
F_E = \langle \psi |\rho_E| \psi \rangle = \sum^{2^N-1}_{\overline{m},\overline{n}=0}|a_{\overline{m},\overline{n}}|^2|\langle \psi| U_{\overline{m},\overline{n}}|\psi\rangle|^2
\label{fidelity}
\end{equation}
and similarly for the $B$ clone (replace the $|a_{\overline{m},\overline{n}}|^2$ term by $|b_{\overline{m},\overline{n}}|^2$).

\section{BB84 protocol with 2-qubit correlated bases}
In this section we compare the amount of information that can be gained by Eve 
when performing a cloning attack on individual qubits (two-dimensional) and 
on pairs of qubits (four-dimensional) which may have been chosen from 
correlated bases.  
We study here
how this affects the BB84 protocol and in the next section
we move on to the six-state protocol.

In the BB84 protocol, Alice chooses from states belonging to two mutually 
unbiased bases. Two bases $A$ and $B$ for a $d$-dimensional system
are said to be MU \cite{LBZ} if a state 
prepared in any element of $A$ (such as $|A,\alpha\rangle$) has a uniform probability distribution of being found in any element of B, namely
\begin{equation}
|\langle A,\alpha | B, \beta\rangle| = \frac{1}{\sqrt{d}}\;.
\end{equation}
Conventionally, Alice and Bob choose the first basis as the so-called computational basis (eigenstates of $\sigma_z$) $\{|0\rangle, |1\rangle\}$ and the second as the dual basis (eigenstates of $\sigma_x$) $\{\frac{1}{\sqrt{2}}(|0\rangle\pm|1\rangle)\}$.  

\subsection{BB84 - single qubit attack - no basis correlation}
If Eve chooses to clone the qubits individually, she must use a cloning strategy which is optimal for this set of states.  
When using the cloning formalism described in Sec. I, one can easily verify that the expression of the fidelity for all states of a given basis is the same. The reader familiar with this calculation can easily skip to the next subsection without any loss of generality. 
Here and throughout the paper, we consider fidelities as expressed by Eq. 
(\ref{fidelity}). Particularly for Eve's clone one finds that the fidelity for the computational basis is $F_E = |a_{0,0}|^2 + |a_{0,1}|^2$ and the dual basis is $F_E = |a_{0,0}|^2 + |a_{1,0}|^2$.  A cloning machine that acts equally well for this set of states implies $|a_{0,0}|^2 + |a_{0,1}|^2 = |a_{0,0}|^2 + |a_{1,0}|^2$.        
Since there is $a$ $priori$ no reason why the optimal values of these elements be different from each other, we make the hypothesis that they should all be equal and real.  Furthermore, we extend our hypothesis to the remaining element, $|a_{1,1}|^2$ such that the form of the amplitude matrix reduces to: 
\begin{eqnarray}
a_{\overline{m},\overline{n}} =\left ( \begin{array}{cc}
v & x\\
x & y
\end{array} \right). \label{amnmatrix}
\end{eqnarray}
Eve's fidelity is now expressed as $F_E = v^2 + x^2$ and normalization requires $v^2 + 2x^2 + y^2 = 1$.  Bob's clone can be characterized by a similar amplitude matrix by making the same hypotheses:
\begin{eqnarray}
b_{\overline{m},\overline{n}} =\left ( \begin{array}{cc}
v' & x'\\
x' & y' 
\end{array} \right),
\end{eqnarray}
where the different matrix elements are related to the $a_{m,n}$ coefficients by Eq.~(\ref{ft}). Thus, Bob's fidelity is $F_B = v'^2 + x'^2$ in both bases and the corresponding mutual information between Alice and Bob (if the latter measures his clone in the good basis) is given by 
\begin{equation}
I_{AB} = 1 + F_B\log_2F_B + (1-F_B)\log_2(1-F_B).
\label{info}
\end{equation}
Maximizing Eve's fidelity $F_E$ for a given value of Bob's fidelity $F_B$ under the normalization constraint yields
\begin{eqnarray}
v&=&\frac{1}{2}+\sqrt{F_B(1-F_B)} \nonumber \\
x&=&F_B-\frac{1}{2}\nonumber \\
y&=& \frac{1}{2}-\sqrt{F_B(1-F_B)}  \nonumber
\end{eqnarray}
such that the corresponding optimal fidelity for Eve is
\begin{equation}
F_E = \frac{F_B}{2}+\frac{1-F_B}{2} + \sqrt{F_B(1-F_B)}.
\end{equation}
Csisz\'{a}r and K\"{o}rner's theorem \cite{CK} provides a lower bound on the rate $R$ at which Alice and Bob can generate secret key bits using privacy amplification:
\begin{equation}
R \ge \mbox{max}(I_{AB}-I_{AE},I_{AB}-I_{BE})\;,
\end{equation}
where $I_{AE}$ and $I_{BE}$ represent the mutual information between Alice and 
Eve, and Bob and Eve respectively.  
It is therefore a sufficient condition that $I_{AB}>I_{AE}$ in order to establish a secret key with non-zero rate for one way communication channels.
It has been shown in \cite{CBKG} that Bob and Eve's information curves 
intersect exactly where the fidelities coincide because, in this particular case, the mutual information shared between Alice and Eve is also expressed by Eq. (\ref{info}).  This yields the optimal symmetric fidelity of phase covariant 
cloning \cite{BCAM}
\begin{equation}
F_E = F_B = \frac{1}{2}+\frac{1}{\sqrt{8}} \simeq 0.8536.
\end{equation}
Note that this result is independent of the fact that Alice may have chosen to 
encode sequences of consecutive qubits in the same basis since Eve is 
intercepting them individually.       

\subsection{BB84 - two qubit attack - no correlation}
Suppose now that Eve intercepts the qubits in sequences of two and clones
them. We make the same assumption as before, namely that Alice has
randomly chosen the basis she has encoded her qubit with.  We would like
to know if Eve can gain more information per qubit using this cloning
approach as opposed to cloning them individually.  Our first task is to
determine the set of states that she will have to clone.  If Alice chooses
among the computational and dual bases, the possible sequences Eve might
encounter are products of eigenstates of $\sigma_z^{\otimes 2}$:
$|00\rangle, |01\rangle, |10\rangle, |11\rangle$, products of eigenstates
of $\sigma_x^{\otimes 2}$:
\begin{eqnarray}
&&\frac{1}{2}(|00\rangle + |01\rangle + |10\rangle + |11\rangle ), \nonumber \\
&&\frac{1}{2}(|00\rangle - |01\rangle + |10\rangle - |11\rangle ), \nonumber \\
&&\frac{1}{2}(|00\rangle + |01\rangle - |10\rangle - |11\rangle ), \nonumber \\
&&\frac{1}{2}(|00\rangle - |01\rangle - |10\rangle +|11\rangle ), \nonumber 
\end{eqnarray}
and products between eigenstates of these two bases
($\sigma_z\otimes\sigma_x$ and $\sigma_x\otimes\sigma_z$):
\begin{eqnarray}
&&\frac{1}{\sqrt{2}}(|00\rangle\pm|01\rangle) 
~~~,~~~\frac{1}{\sqrt{2}}(|10\rangle\pm|11\rangle) \nonumber \\
&&\frac{1}{\sqrt{2}}(|00\rangle\pm|10\rangle)~~~,~~~\frac{1}{\sqrt{2}}(|01\rangle\pm|11\rangle). 
\nonumber
\end{eqnarray}
Because we are now dealing with a $four$-dimensional Hilbert space ($N=2$)
with tensor product structure, the $U_{\overline{m},\overline{n}}$
operators take the following form:
\begin{eqnarray}
U_{m_{1}m_{2};n_{1}n_{2}}=
\left ( \begin{array}{cc}
I & Z\\
X & Y\\
\end{array} \right) \otimes
\left ( \begin{array}{cc}
I & Z\\
X & Y\\
\end{array} \right) 
\nonumber
\end{eqnarray}
Each of these matrix elements consists in a tensor product of two Pauli
operators each acting on an associated qubit.  Eve is interested in the
information she can gain from a single qubit when she clones them in
sequences of two.  In other words, Eve is interested in the optimal
$four$-dimensional cloning map where the figure of merit is not the
single-clone $four$-dimensional fidelity but rather the single-clone,
single-qubit $two$-dimensional fidelity averaged over the two qubits. To
obtain this fidelity, we must trace over the second qubit subsystem and
compute the fidelity of the first qubit, repeat this operation for the
second qubit by tracing out the first qubit subsystem and finally average
over the two fidelities.  For example, the reduced density matrix of the
first qubit for Eve's clone is expressed as:
\begin{eqnarray}
\rho_E^1 &=& \mbox{Tr}_2 \big[\ \sum_{\overline{m},\overline{n}}|a_{\overline{m},\overline{n}}|^2X^{m_1}Z^{n_1}|\phi_1\rangle\langle \phi_1|Z^{n_1}X^{m_1} \nonumber \\
&\otimes& X^{m_2}Z^{n_2}|\phi_2\rangle\langle \phi_2|Z^{n_2}X^{m_2}\big] \nonumber \\
&=&
\sum_{\overline{m},\overline{n}}|a_{\overline{m},\overline{n}}|^2X^{m_1}Z^{n_1}|\phi_1\rangle\langle \phi_1|Z^{n_1}X^{m_1} \nonumber 
\end{eqnarray} 
where $|\phi_i\rangle$ is a two-dimensional system.
For sequences of qubits both drawn from eigenstates of $\sigma_z$ the
fidelity is
\begin{eqnarray}
F_{E,zz}^1 &=& \sum_{\overline{m}=0,\overline{n}=0}^{2^N-1}|a_{\overline{m},\overline{n}}|^2| \langle \phi_1|X^{m_1}Z^{n_1}|\phi_1\rangle|^2  \nonumber \\
&=& \sum_{\overline{m}=0,\overline{n}=0}^{2^N-1} |a_{\overline{m},\overline{n}}|^2 \delta_{m_1,0} 
\label{z1}
\end{eqnarray}
for the first qubit and 
\begin{eqnarray}
F_{E,zz}^2 &=& \sum_{\overline{m}=0,\overline{n}=0}^{2^N-1}|a_{\overline{m},\overline{n}}|^2| \langle \phi_2|X^{m_2}Z^{n_2}|\phi_2\rangle|^2 \nonumber \\
&=& \sum_{\overline{m}=0,\overline{n}=0}^{2^N-1} |a_{\overline{m},\overline{n}}|^2 \delta_{m_2,0}	
\label{z2}
\end{eqnarray}
for the second qubit.
For clusters of qubits both drawn from eigenstates of $\sigma_x$ the
fidelity is
\begin{eqnarray}
F_{E,xx}^1 &=& \sum_{\overline{m}=0,\overline{n}=0}^{2^N-1} |a_{\overline{m},\overline{n}}|^2 \delta_{n_1,0} 
\label{x1}
\end{eqnarray}
for the first qubit
and
\begin{eqnarray}
F_{E,xx}^2 &=& \sum_{\overline{m}=0,\overline{n}=0}^{2^N-1} |a_{\overline{m},\overline{n}}|^2 \delta_{n_2,0} 
\label{x2}
\end{eqnarray}
for the second qubit.
To be complete, we must also compute the fidelity for clusters expressed as tensor products drawn from eigenstates of $\sigma_z\otimes\sigma_x$ and $\sigma_x\otimes\sigma_z$.  The former yields 
$F_{E,zx}^1 = F_{E,zz}^1$ for the first qubit
and $F_{E,zx}^2 = F_{E,xx}^2$ for the second qubit.
The latter yields a fidelity of 
$F_{E,xz}^1 = F_{E,xx}^1$ for the first qubit and $F_{E,xz}^2 = F_{E,zz}^2$ for the second qubit.
The expressions for these fidelities $F_{E}^i$ can easily be interpreted as follows.  
Every single-qubit fidelity consists in a sum of eight terms for which the first four express the fidelity of the $four$-dimensional system in question (in other words the contribution from the $a_{\overline{m},\overline{n}}$ coefficients where no errors occur on either qubits) while the remaining four terms correspond to the
$a_{\overline{m},\overline{n}}$ coefficients for which the $i^{\mbox{th}}$
qubit is not affected by an error but the remaining one is.
Generally, the fidelity of the $i^{\mbox{th}}$ qubit is expressed as 
\begin{eqnarray}
F_E^i = F_{4E} + D_E^i
\end{eqnarray}
where $F_{4E}$ is the fidelity of the $four$-dimensional system and $D_E^i$ is the disturbance of the $i^{\mbox{th}}$ qubit and is expressed as
\begin{equation}
D_E^i = \sum_{\overline{m}=0,\overline{n}=0}^{2^N-1}|a_{\overline{m},\overline{n}}|^2 \delta_{m_i,1} \, \delta_{m_{\neg i},0}
\end{equation} 
for qubits drawn from the computational basis and 
\begin{equation}
D_E^i = \sum_{\overline{m}=0,\overline{n}=0}^{2^N-1}|a_{\overline{m},\overline{n}}|^2
\delta_{n_i,1} \, \delta_{n_{\neg i},0}
\end{equation}
for qubits drawn from the dual basis.
Here, the qubit of the pair which is not the $i^{\mbox{th}}$ qubit
is given the index $\neg i$. The average qubit fidelity of Eve's clone is therefore:
\begin{eqnarray}
F_E = F_{4E} + \frac{1}{2}(D_E^1+D_E^2).
\label{F2DE}
\end{eqnarray}
A similar analysis can be made for Bob's clone from which we obtain a single-qubit fidelity
\begin{eqnarray}
F_B = F_{4B} + \frac{1}{2}(D_B^1+D_B^2)
\label{F2DB}
\end{eqnarray}
which is function of the $b_{\overline{m},\overline{n}}$ coefficients.
We are again interested in the mutual information shared between Alice and Bob and Alice and Eve.  To do this, let us first compute Eve's optimal fidelity $F_E$ for a fixed value of Bob's fidelity $F_B$ under the normalization constraint 
\begin{eqnarray}
\sum_{\overline{m}=0,\overline{n}=0}^{3}|a_{\overline{m},\overline{n}}|^2 = 1 \label{norm}
\end{eqnarray}
and the constraint that the single-qubit fidelity be the same for all 16 considered input states.  The optimization yields the following $a_{\overline{m},\overline{n}}$ matrix:
 \begin{eqnarray}
a_{\overline{m},\overline{n}} 
&=& \left ( \begin{array}{cc}
v_1 & x_1\\
x_1 & y_1
\end{array} \right)
\otimes 
\left ( \begin{array}{cc}
v_2 & x_2\\
x_2 & y_2
\end{array} \right) \label{amn}
\end{eqnarray}
where
\begin{eqnarray}
v_1 &=& v_2 = \frac{1}{2}+\sqrt{F_B(1-F_B)} \nonumber \\
x_1 &=& x_2 = F_B-\frac{1}{2}\nonumber \\
y_1 &=& y_2 = \frac{1}{2}-\sqrt{F_B(1-F_B)}  \nonumber
\end{eqnarray}
such that 
\begin{equation}
F_E = \frac{F_B}{2}+\frac{1-F_B}{2} + \sqrt{F_B(1-F_B)}.
\label{f84}
\end{equation}
From the previous subsection we know that 
Bob and Eve's information curves intersect exactly where the fidelities 
coincide. This implies that Alice and Bob can share secret bits via privacy 
amplification as long as $F_B > F_E$, that is
\begin{equation}
F_B > \frac{1}{2} + \frac{1}{\sqrt{8}}.\nonumber
\end{equation}
This optimal symmetric fidelity turns out to be the same as the optimal fidelity obtained when the cloner is designed for $two$-dimensional systems meaning that the optimal $four$-dimensional cloning map for single-qubit single-clone fidelity boils down to the tensor product of the $two$-dimensional optimal cloners.        
\subsection{BB84 - two qubit attack - correlated bases}
Now consider the situation where Alice is limited by her random number generator and must therefore send two consecutive states drawn from the same basis in order to keep a decent cadence \cite{gisin}.  Of course if Eve intercepts every qubit individually, the fidelity she obtains after cloning is just the same as before, namely $F = \frac{1}{2} + \frac{1}{\sqrt{8}}$.  If she intercepts them in 
sequences of two qubits she will necessarily find that they are correlated: either
she expects to find two qubits drawn from the computational basis $\sigma_z$ (equivalently, a four dimensional state drawn from the eigenstates of $\sigma_z\otimes\sigma_z$) or two qubits drawn from the dual basis $\sigma_x$ (equivalently, a four dimensional state drawn from the eigenstates of $\sigma_x\otimes\sigma_x$).  Compared to the previous situation where no correlation was present, the set of input states Eve has to consider has now decreased.  Intuitively we should expect that the optimal single-qubit cloner would give rise to a higher fidelity.  We shall see that this is not the case.  

The cloner we consider is again characterized by the "Cerf ansatz" (\ref{ansatz}) such that the single-qubit fidelity for this set of input states is defined exactly like Eqs. (\ref{z1}) and (\ref{z2}) for eigenstates of $\sigma_z$ and like Eqs. (\ref{x1}) and (\ref{x2}) for eigenstates of 
$\sigma_x$.  These are the four expressions of the fidelity for which the $a_{\overline{m},\overline{n}}$ (and consequently the $b_{\overline{m},\overline{n}}$) coefficients must be optimized for.  The constraints we must consider here are the normalization constraint and the constraint that these four expressions be equal.  Of course, these fidelities are again characterized by Eq. (\ref{F2DE}). Interestingly, the constrained optimization yields $a_{\overline{m},\overline{n}}$ coefficients which have exactly the same form as Eq. (\ref{amn}) and therefore the same expressions for Eve's fidelity as a function of Bob's.  Once again, the lower bound on the mutual information Alice and Bob must share in order to generate a secret key is given by
\begin{equation}
F > \frac{1}{2} + \frac{1}{\sqrt{8}}. \nonumber
\end{equation} 
We conclude that even if Alice chooses to encode two consecutive states in the same basis, Eve's optimal cloning strategy does not permit her to gain more information than complete random choices.  
In Section V we will generalize this idea for sequences of $N$ qubits, 
but first let us examine how these cloning strategies apply to the six-state 
protocol.

\section{Six-state protocol with 2-qubit correlated bases}

The six-state protocol is very similar to the BB84 protocol, the only difference being that Alice now has the 
choice to pick up states from a third basis MU to the other two.  
Again, let us choose the first two bases as 
the computational basis and the dual basis and let the third basis be the eigenstates of $\sigma_y$: 
$\{\frac{1}{\sqrt{2}}(|0 \rangle \pm i|1 \rangle)\}$.  

\subsection{six-state - single qubit attack - no correlation}

The cloner that must be used for the six-state protocol is an asymmetric 
$two$-dimensional universal cloner \cite{CBKG} characterized by the same amplitude matrix as Eq. 
(\ref{amnmatrix}) except that we make the change $y = x$: \begin{eqnarray}
a_{\overline{m},\overline{n}} &=&\left ( \begin{array}{cc}
v & x\\
x & x 
\end{array} \right) \nonumber.
\end{eqnarray}
Eve's fidelity is expressed as $F_E = v^2 + x^2$ and normalization requires $v^2 + 3x^2 = 1$.  Maximizing her fidelity for a fixed value of Bob's fidelity yields the optimal cloner:
\begin{eqnarray}
v &=& \sqrt{\frac{3F_B-1}{2}} \nonumber \\
x &=& \sqrt{\frac{1-F_B}{2}} \nonumber.
\end{eqnarray}
Bob's clone is characterized by a similar amplitude matrix:
\begin{eqnarray}
b_{\overline{m},\overline{n}} =\left ( \begin{array}{cc}
v' & x'\\
x' & x' 
\end{array} \right),
\end{eqnarray}
where as before, $v'$ and $x'$ are given by Eq. (\ref{ft}) while the mutual information he shares with Alice by Eq. (\ref{info}).  It has been shown in 
\cite{CBKG} that the mutual information shared between Alice and Eve for the six-state protocol is given by 
\begin{eqnarray}
I_{AE} &=& 1+(F_B+F_E-1)\log_2(\frac{F_B+F_E-1}{F_B})\nonumber \\
&+& (1-F_E)\log_2(\frac{1-F_E}{F_B})
\end{eqnarray}  
such that for a given $F_B$, $I_{AE}$ is lower than for the BB84 protocol which is consistent with the stronger requirement we put on that cloner.  This implies that the fidelity $F_B$ for which $I_{AE} = I_{AB}$ is slightly lower, 
and equal to $F_B \simeq 0,8436$. 

\subsection{six-state - two qubit attack - no basis correlation}
If Eve chooses to clone the incoming states in sequences of two,  
the set of four-dimensional states she has to clone consists of tensor products of states belonging to the three maximally unbiased bases above. The 
single-qubit fidelity is computed as above, 
with the exception that there are extra constraints, namely that the fidelity should also clone equally well eigenstates of $\sigma_y$:
\begin{eqnarray}
F_{E,yy}^1 &=& \sum_{\overline{m}=0,\overline{n}=0}^{2^N-1} |a_{\overline{m},\overline{n}}|^2 \delta_{m_1,n_1}  \nonumber
\label{y1}
\end{eqnarray}
for the first qubit and 
\begin{eqnarray}
F_{E,yy}^2 &=& \sum_{\overline{m}=0,\overline{n}=0}^{2^N-1} |a_{\overline{m},\overline{n}}|^2 \delta_{m_2,n_2} \nonumber
\label{y2}
\end{eqnarray}
for the second qubit.
The other constraints come from tensor products of $\sigma_y\otimes\sigma_z$, $\sigma_y\otimes\sigma_x$ and vice-versa.  
The expression for the fidelity of the $i^{\mbox{th}}$ qubit can be expressed as:
\begin{eqnarray}
F_{E}^i = F_{4E} + D_E^i 
\end{eqnarray}
where, for eigenstates of $\sigma_y$,
\begin{eqnarray}
D_E^i = \sum_{\overline{m}=0,\overline{n}=0}^{2^N-1}|a_{\overline{m},\overline{n}}|^2
\delta_{m_{i},n_{i}+1} \, \delta_{m_{\neg i},n_{\neg i}} .
\end{eqnarray}
The average qubit fidelity is again:
\begin{eqnarray}
F_E = F_{4E} + \frac{1}{2}(D_E^1+D_E^2).
\end{eqnarray}
As before a similar analysis can be made for Bob's clone from which we obtain a single-qubit fidelity
\begin{eqnarray}
F_B = F_{4B} + \frac{1}{2}(D_B^1+D_B^2).
\end{eqnarray}
We are again interested in the mutual information shared between Alice and Bob,
and Alice and Eve.  We compute Eve's optimal fidelity $F_E$ for a fixed value of Bob's fidelity $F_B$ under the normalization constraint Eq.(\ref{norm}) and the constraint that the single-qubit fidelity be the same for all input states.  The optimization yields the following $a_{\overline{m},\overline{n}}$ matrix:
 \begin{eqnarray}
a_{\overline{m},\overline{n}}
&=& \left ( \begin{array}{cc}
v_1 & x_1\\
x_1 & x_1
\end{array} \right)
\otimes 
\left ( \begin{array}{cc}
v_2 & x_2\\
x_2 & x_2
\end{array} \right) 
\end{eqnarray}
where
\begin{eqnarray}
v_1 &=& v_2 =  \sqrt{\frac{3F_B-1}{2}} \nonumber \\
x_1 &=& x_2 = \sqrt{\frac{1-F_B}{2}}\nonumber 
\end{eqnarray}
such that 
\begin{equation}
F_E = 1-\frac{F_B}{2} + \frac{1}{4}\sqrt{6F_B-2}\sqrt{2-2F_B}.
\label{f6}
\end{equation}
In the previous subsection, we have seen how to express 
$I_{AB}$ and $I_{AE}$. 
Again in this case the lower bound on Bob's fidelity needed for $I_{AB} > I_{AE}$ is given by $F_B > 0.8436$ which is the same fidelity for individual attacks.
Thus, so far, we arrive to the same conclusions as for the BB84 protocol.  
\subsection{six-state - two qubit attack - correlated bases}
If Alice is again limited by her random number generator and must encode two consecutive qubits in the same basis, Eve can clone the incoming states by 
sequences of two expecting to find four-dimensional states expressed as eigenstates of $\sigma_z\otimes 
\sigma_z$,  $\sigma_x\otimes\sigma_x$ or $\sigma_y\otimes\sigma_y$.
By making a similar reasoning as in the previous subsection we arrive to the same conclusions as before, namely that the information Eve can gain when cloning a $four$-dimensional system boils down to the optimal single qubit information.

\section{Cloning of $N$-qubit sequences}
We now proceed to generalize the cloning strategies considered in the previous sections.  We suppose that Alice encodes her qubits using the same basis for 
sequences of $N$ qubits.  We also suppose that $N$ is much smaller than the total size of the raw key she will be exchanging with Bob.  
We also suppose that Eve is aware of when a new sequence begins and ends. 

Generally, for a sequence of $N$ qubits, the reduced density matrix of the $i^{\mbox{th}}$ qubit for a given clone (say $E$) is written as
\begin{eqnarray}
\rho_E^i &=& \mbox{Tr}_{j\ne i}  \sum_{\overline{m},\overline{n}}|a_{\overline{m},\overline{n}}|^2\bigotimes_{j=1}^N X^{m_j}Z^{n_j}|\phi_j\rangle\langle \phi_j|Z^{n_j}X^{m_j}\nonumber \\
&=& 
\sum_{\overline{m},\overline{n}}|a_{\overline{m},\overline{n}}|^2 X^{m_i}Z^{n_i}|\phi_i\rangle\langle \phi_i|Z^{n_i}X^{m_i}, 
\end{eqnarray}
such that fidelity of the $j^{\mbox{th}}$ qubit is written as
\begin{equation}
F_E^j = F_{E2^N} + D_E^j
\end{equation}
and similarly for qubits of Bob's clone. 
The average qubit fidelity is therefore expressed as:
\begin{equation}
F_E = F_{E2^N} + \frac{1}{N}\sum_{i=1}^{N} D_E^i.
\end{equation}
If we assume that the optimal $a_{\overline{m},\overline{n}}$
amplitude matrices are expressed as 
\begin{eqnarray}
a_{\overline{m},\overline{n}} =\left ( \begin{array}{cc}
v & x\\
x & y
\end{array} \right)^{\otimes N}
\end{eqnarray}  
for the BB84 protocol and 
\begin{eqnarray}
a_{\overline{m},\overline{n}} =\left ( \begin{array}{cc}
v & x\\
x & x
\end{array} \right)^{\otimes N}
\end{eqnarray} 
for the six-state protocol, we can check that they indeed satisfy a
constrained optimization.  Since the information curves are both monotonically increasing functions of the 
fidelities, we use the Lagrange multiplier method to
optimize Eve's fidelity for a fixed value of Bob's.

The constraint that the fidelity for different qubits in the 
sequence be the same is already satisfied by the hypothesized 
$a_{\overline{m},\overline{n}}$ matrix.  The function is:
\begin{widetext}
\begin{eqnarray}
\mathcal{L} &=& F_E + \lambda_1F_B+\lambda_2(\sum_{\overline{m},\overline{n}=0}^{2^N-1}|a_{\overline{m},\overline{n}}|^2-1) \nonumber \\
&=&   \frac{1}{N}\sum_{\overline{m}=0}^N (N-\sum_{i=1}^N m_i)\prod_{i=1}^{N}(v^2+x^2)^{m_i\oplus1}(x^2+y^2)^{m_i} \nonumber \\
&+& \lambda_1\big[
\frac{1}{N}\sum_{\overline{m}=0}^N (N-\sum_{i=1}^N m_i)\prod_{i=1}^{N}
(\frac{1}{2}+vx+xy)^{m_i\oplus1}(\frac{1}{2}-vx-xy)^{m_i}\big]\nonumber \\
&+& \lambda_2 \big[ (v^2+2x^2+y^2)^N -1 \big]
\label{eq:La}
\end{eqnarray}
\end{widetext}
where the modular sum is in base 2.  The equivalent expression of Eq. (\ref{eq:La}) for the six-state protocol 
is 
very 
similar except that one should exchange $y^2$ for
$x^2$. 

We have checked, using a symbolic calculator, that the hypothesized amplitude matrices satisfy the
constrained optimization and yield the optimal fidelities (Eqs. (\ref{f84})
and (\ref{f6})) for $N=2$ and $N=3$.

\section{Conclusion}
We have considered the cloning of sequences of $N$ qubits, where 
all the qubits in each sequence are prepared in the same basis while
each state is chosen at random. This situation is 
very different from the usual scenario of cloning multiple copies, 
where all the copies are prepared in the same state. Our investigation was 
motivated by the situation in quantum cryptography where the legitimate 
users are required to make truly random choices for each single qubit. 
From a practical point of view, this requirement on high-speed 
random-number generation is a severe constraint.  

However, under the assumption that the class of cloning transformations 
we considered here provides the optimal eavesdropping strategy, 
we have shown that this requirement can be relaxed, 
so that Alice can prepare long sequences of qubits in 
the same basis without compromising the security. 
Surprisingly, Eve cannot exploit her knowledge that the used basis
is fixed for the entire sequence, regardless of its length
(provided it is much shorter that the total key size). This result
is quite important for practical applications of quantum cryptography
as it implies that higher secret-key rates may be obtained using the same
random number generator but with this new modified protocol.

\section*{Acknowledgment}
The authors acknowledge financial support from the Action de
Recherche Concert{\'e}e de la Communaut\'e Fran{\c{c}}aise de Belgique,
from the IUAP program of the Belgian Federal Governement under grant V-18
and from the European Union through projects RESQ (IST-2001-37559) and
SECOQC (IST-2003-506813).



\begin{thebibliography}{20}
\bibitem{BB84}
C.H. Bennett and G. Brassard, in Proceedings of the IEEE International Conference on Computers, Systems and Signal Processing, Bangalore, India, 1984 (unpublished), p. 175-179.

\bibitem{Ekert}
A. Ekert, Phys. Rev. Lett. \textbf{678}, 661 (1991).

\bibitem{GRTZ} N. Gisin, G. Ribordy, W. Tittel, and H. Zbinden, Rev. Mod. Phys. \textbf{74}, 145, 2002.


\bibitem{WZ}
W.K. Wootters and W.H. Zurek, Nature (London) 299, 802 (1982).



\bibitem{dagmar}  D. Bru{\ss}, Phys. Rev. Lett. {\bf 81}, 3018 (1998).

\bibitem{BG}  H. Bechmann-Pasquinucci and N. Gisin, 
           Phys. Rev. A {\bf 59}, 4238 (1999).

\bibitem{griffiths}  C.~Fuchs, N.~Gisin, R.~Griffiths, 
C.-S.~Niu and A.~Peres, Phys. Rev. A {\bf 56}, 1163 (1997).

\bibitem{CBKG} N. J. Cerf, M. Bourennane, A. Karlsson and N. Gisin, 
Phys. Rev. Lett. \textbf{88} 127902 (2002).

\bibitem{gisin} N. Gisin, quant-ph/0303052 (2003).


\bibitem{cerfansatz1} N. J. Cerf, Proc. 1st NASA International Conference
QCQC'98, Palm Springs, February 1998; 
also in Acta Phys. Slov. {\bf 48}, 115 (1998).

\bibitem{cerfansatz2} N. J. Cerf, Phys. Rev. Lett. {\bf 84}, 4497 (2000);
  J.  Mod. Opt. {\bf 47}, 187 (2000).

\bibitem{LBZ} J. Lawrence, C. Brukner and A. Zeilinger Phys. Rev. A \textbf{65}, 032320 (2002)

\bibitem{CK} I. Csisz\'{a}r and J. K\"{o}rner, IEEE Trans. Inf. Theory \textbf{24}, 339 (1978).





\bibitem{BCAM} D. Bru\ss, M. Cinchetti, G. M. D Ariano, and C. Macchi- avello, Phys. Rev. A \textbf{62} 012302 (2000).






\end{thebibliography}
\end{document}